\documentclass[twocolumn,prd,aps,amsmath,amssymb,nofootinbib]{revtex4-2}
\usepackage{graphics}
\usepackage{graphicx}
\usepackage{dcolumn}
\usepackage{bm}
\usepackage{mathrsfs}
\usepackage{pstricks}
\usepackage{color}
\usepackage{slashed}
\usepackage{amsmath}
\usepackage{epsfig}
\usepackage{amsfonts}
\usepackage{amssymb}
\def\beq{\begin{equation}}
\def\eeq{\end{equation}}
\def\bea{\begin{eqnarray}}
\def\eea{\end{eqnarray}}
\def\nn{\nonumber}
\def\lan{\langle}
\def\ran{\rangle}
\def\r{\mathbf{r}}
\def\Re{\textrm{Re}}
\def\Im{\textrm{Im}}

\def\x{\mathbf{x}}
\def\y{\mathbf{y}}

\def\p{\mathbf{p}}
\RequirePackage[normalem]{ulem}
\RequirePackage{color}

\begin{document}

\title{Structural glasses model using disorder fields: \\the boson peak from local ground states}

\author{M. M. Balbino}
\email{matheusdemb@cbpf.br}
\affiliation{Centro Brasileiro de Pesquisas F\'{\i}sicas, 22290-180 Rio de Janeiro, RJ, Brazil}

\author{I. P. de Freitas}
\email{isaquepfreitas@cbpf.br}
\affiliation{Centro Brasileiro de Pesquisas F\'{\i}sicas, 22290-180 Rio de Janeiro, RJ, Brazil}

\author{N.~F.~Svaiter}
\email{nfuxsvai@cbpf.br}
\affiliation{Centro Brasileiro de Pesquisas F\'{\i}sicas, 22290-180 Rio de Janeiro, RJ, Brazil}

\author{A. M. S. Macedo}
\email{antoniosmacedo@ufpe.br}
\affiliation{Universidade Federal de Pernambuco - UFPE, Recife, PE, Brazil}

\author{G.~Krein}
\email{gastao.krein@unesp.br}
\affiliation{Instituto de F\'{i}sica Te\'orica, Universidade Estadual Paulista, 01140-070 S\~ao Paulo, SP, Brazil}


\begin{abstract}
We show the emergence of a contribution characteristic of {the} boson peak in the spectral density of structural glasses. To model the vitreous state, we consider static density-fluctuation fields coupled to a multiplicative quenched disorder. Performing an ensemble average over all disorder realizations, a functional series representation of the average free energy is obtained. In this series representation of the average free energy for the glassy state of matter, we identify in the function space effective actions. These effective actions present a large number of metastable states and ground states. Random first-order transition, widely discussed in the literature as a description of the transition from the supercooled liquid to the glassy state of matter, emerges naturally in our formalism. {We establish the  connection between the use of hyperbolic differential equations with random coefficients and the presence of many ground states in the average free energy. This connection allows us to study emergent excitations in such amorphous materials.} 
\end{abstract}



\maketitle

%
\section{Introduction}\label{intro}

Crystallization {constitutes} a first-order phase transition between two thermodynamically stable phases: the liquid phase and the crystalline solid phase. {This transition} is typically accompanied by {an abrupt decrease in} specific volume. {The resulting} crystalline solids are symmetry-broken states characterized by long-range order. {In this phase, ergodicity} is broken, {since the} disordered configurations of the liquid state are no longer accessible to the system. {In the liquid, upon a rapid} temperature quenching, when the cooling rate exceeds the structural {relaxation} rate, one {instead} observes a {pronounced increase} in viscosity. Below the melting/crystallization temperature, the system becomes a glass, a material with an extremely large viscosity and non-zero shear moduli \cite{binder}.

Structural glasses are systems that exhibit mechanical rigidity comparable to that of crystalline materials; however, they lack long-range structural order as amorphous solids. At the molecular level, the {structural organization} of a glass is very similar to that of a liquid~\cite{Takeno,Brazhkin,Tanaka2}, {despite their markedly different} dynamical properties. {In the glassy phase, ergodicity} is also broken {because the dynamics exhibit a pronounced} slowdown, {so that the system is no longer able} to explore its configuration space {within} experimental time scales. The {central challenge  for developing a comprehensive} understanding of the non-ergodic glassy phase {lies in rationalizing the apparent decoupling between structural characteristics and dynamical behavior.} {Although the dynamical slowdown is expected to originate at the microscopic level, the literature has different proposals to its underlying mechanism.}

There are two main approaches to modeling vitrification without crystallization \cite{Jackle,Debe,Dyre,cavagna,Biroli1,Garrahan,royall,Loidl}. One {approach} invokes thermodynamics, {whereas} the other claims that the transition from a supercooled liquid to the glassy state is {purely dynamical}. The {thermodynamic framework} is based on an equilibrium theory using density functional theory,  which was developed to study the ground state properties of many-body systems in terms of expectation values of particle-density operators \cite{kohn}. The foundation of this formalism is the idea that the free energy is a functional of some a coarse-grained density field, which attains its minimum at the equilibrium state. For conventional systems, the most stable structure {can be identified} by comparing the free energy {values associated with different candidate } structures. This static description of the structural glass transition uses a free energy functional with replicas~\cite{re,emery}. The second {approach is dynamical mode-coupling} theory, {which aims to capture} the full-scale microscopic physics of glass-forming liquids. {In this approach, one analyses} the time-dependent density correlation functions of the liquid \cite{gotze22,janssen,das}. Within a mean-field approximation, this dynamic mode-coupling theory predicts an ergodicity-breaking transition accompanied by a growing dynamical length as a critical point is approached. However, as a perturbative approach, it is not able to access nonperturbative collective phenomena. This is an important issue, since there is {a broad consensus} within the condensed-matter community that nonperturbative effects {plays} an important role in the glass transition.

Glasses exhibit low-temperature properties that differ markedly from those of crystalline materials~\cite{Pohl,phillips}. At very low temperatures, the specific heat of glasses grows approximately as $T$ from zero temperature, whereas the thermal conductivity grows approximately as $T^{2}$. This behavior differs substantially from that predicted by the Debye model for crystals. The anomalous temperature dependence of the specific heat appears in different systems, but these effects are accentuated in strong liquids at low temperatures. This {terminology ``strong'' and ``fragile'' arises} because not all systems vitrify in the same manner. {The} differences {in their vitrification behavior} are characterized by the slope of the viscosity as a function of temperature near the glassy phase and are referred to as fragility. Strong liquids exhibit Arrhenius-type growth of the viscosity, i.e., a smooth growth of the relaxation time, whereas fragile liquids vitrify abruptly with super-Arrhenius behavior; thus, the thermodynamic properties of a glassy state depend on how it was formed.

Furthermore, the lack of periodicity raises fundamental questions concerning the spectrum of elementary excitations in the vitreous state of matter. In the Debye theory of crystalline solids, the density of states follows $D(\omega)=A_{D}\,\omega^{2}$, where $\omega$ is the angular frequency and $A_{D}$ depends on the elastic properties of the material but is independent of frequency \cite{Debye,Peter}. 
Amorphous materials exhibit the boson peak, an excess vibrational density of states over that predicted by the phonon Debye model \cite{Ganter, gurevich, gurevich2,
Vacher, shint, wang, tanaka}. In non-crystalline solids, this excess appears as a peak in $D(\omega)/\omega^{2}$ at an intermediate frequency. {The study of elementary excitations in structurally disordered materials is particularly challenging because the usual methods based on plane waves and phonons are not directly applicable. Therefore, a} major difficulty in developing an analytic theory for the spectrum of elementary excitations in the glassy phase is the presence of multiple metastable and local ground states in the free energy landscape.

{The primary objective of this work is to {discuss quasi-localized excitations in the glassy state of matter by employing} a continuous field-theoretic framework, formally analogous to a Euclidean quantum field theory in $\mathbb{R}^d$. A field-theoretic treatment of}
amorphous solids {based on} a different scenario was discussed in Refs.~\cite{De1,De2}. {Here, we use an approach based on a continuous Ising-like model~\cite{Langer2}.} It is important to emphasize that a {relation} between the multiplicity of local ground states in the average free energy of a glassy phase and the emergence of the boson peak has been proposed in Ref.~\cite{boson}. By adopting a distinct approach, we likewise demonstrate a connection between the presence of many local minima in the free energy landscape and the boson peak. Since the formation of a non-crystalline solid can occur in any viscous liquid with very different microscopic structures, provided that it is cooled rapidly enough, we use a coarse-grained description of the microscopic {degrees of freedom} of the amorphous system. Therefore, even in the absence of an order parameter, {we invoke} the {notion} of collective behavior {and argue that} a continuous field theory {offers} a useful framework for describing the {glassy state of matter}.
{We first formulate} a field theory for fluctuations in the glassy phase {based on} coarse-grained fields {subjected to quenched} randomness. {In continuous field-theory approaches to critical phenomena, coarse-grained fields are defined by averaging over domains that are large compared with the microscopic scale but small compared with the characteristic size of critical fluctuations, i.e., the correlation length of the model. In second-order phase transitions, the correlation length diverges. This is not the situation considered here. Instead, the size of the dynamically heterogeneous domains defines the relevant length scale of the system.} 

An unresolved debate in the literature concerns whether the {glassy phase} is more appropriately described {by Potts-glass-type models}~\cite{kir2}, random-field scenarios~\cite{stevenson,biroli,rizzo1}, or spin-glass physics~\cite{moore}. {In the context of glassy phases, the concept of dynamic heterogeneity has been extensively employed to characterize their dynamical behavior, based on the observation that structural relaxation does not proceed uniformly throughout the sample volume~\cite{sillescu,ediger}.} The {associated} environmental randomness, {commonly referred to in} the literature {as} self-induced disorder, {motivates} the introduction of quenched disorder in models of supercooled liquids~\cite{quenched}. {We use the results discussed in Ref.~\cite{Tanaka}, where heterogeneous dynamics arises from static fluctuations in the average density of the glass.} Here, we consider static density-fluctuation fields coupled to multiplicative quenched disorder, {in direct analogy with} a continuous Landau-Ginzburg model with a random temperature parameter ({in} the terminology of Ref.~\cite{sherin}). We define a coarse-grained static density-fluctuation field using a field theory in $\mathbb{R}^d$ with a formalism similar to Euclidean quantum field theory, where one defines Green's functions in imaginary time, namely the Schwinger functions~\cite{Schwinger, Nagano, Symanzik,Brydges, jaffe}. {Within this formulation, a random first-order transition, which has been widely discussed in the literature 
as a description of the transition from the supercooled liquid to the glassy state of matter~\cite{kir,kir3,xia,Bouchaud}, {emerges} naturally.} Next, we discuss the connection between the presence of many local minima that characterize the average free energy and the use of hyperbolic differential equations with random coefficients to describe emergent excitations. In the scenario of amorphous materials, the {characteristic} boson-peak behavior is derived.  

{To address the problem of characterizing thermodynamic amorphous order and deriving the boson peak, we first compute the disorder-averaged generating functional of connected correlation functions by employing the distributional zeta-function formalism \cite{distributional, distributional2,zarro,robinson,mojica,heymans22, heymans24,freitas, heymans}. This procedure yields a series representation of the disorder-averaged free energy.} 
A particular choice in function space shows that the series representation of the average free energy can model a system trapped in one amorphous state. The resulting free energy landscape contains a large number of distinct local minima with different elastic properties, as commonly expected in structural glasses. {This multivalley structure provides a field-theoretic realization of the random first-order transition scenario for glass formation}. From this landscape, we model the elastic behavior of the glassy state and show how random differential equations can be used to describe quasi-localized excitations in non-crystalline materials.
\cite{sokolov,Gurarie1,Gurarie2,sheng}.  
We describe the emergent excitations in glassy materials, i.e., the non-phononic excitations of the glassy state of matter, and obtain the excess vibrational density of states discussed in the literature. We show the appearance of an $\omega^{4}$ contribution in the spectral density $D(\omega)$ \cite{Hunklinger, Baggioli3, Baggioli1, Lerner}.

The structure of this work is organized as follows. In Sec.~\ref{sec:fe2}, we introduce the field-theoretic framework with multiplicative disorder to model the glass state of matter. In Sec.~\ref{sec:ftg}, we analyze the multivalley structure of the free energy landscape and elucidate the emergence of quasi-localized excitations in structural glasses. In Sec.~\ref{sec:ptdm}, we derive the excess of low-frequency vibrational modes, namely, the vibrational density of states. {The main conclusions are summarized in Sec.~\ref{sec:conclusions}. In Appendix~A, we detail the procedure for mapping the free energy landscape onto a dynamical description of elastic excitations. In Appendix~B, we present the derivation of the spectral density of glassy materials. Throughout this work we adopt natural units, setting $k_{B}= 1$.} 

%
\section{Field theory of fluctuations in glasses}\label{sec:fe2}

{Consider a supercooled liquid whose temperature lies below its freezing point but which has not crystallized.} At the macroscopic level, the fluid dynamics are governed by the Navier-Stokes equation, with bulk and shear viscosity coefficients. By linearizing this equation, invoking an adiabatic assumption, and combining it with the linearized continuity equation, one obtains a lossy acoustic wave equation~\cite{lamb}. In this linear regime, the differential equation describing sound waves in a viscous medium can be written as
\begin{equation}
\frac{\partial^{2}\mathbf{u}_{\lambda}(t,\x)}{\partial t^{2}}=\Delta\biggl(c_{\lambda}^{2}+D_{\lambda}\frac{\partial}{\partial t}\biggr){\mathbf{u}}_{\lambda}(t,\x),
\label{navier}
\end{equation}
where $\Delta$ denotes the Laplacian differential operator in $\mathbb{R}^{d}$. The sound wave $\mathbf{u}_\lambda(t,\x)$ can be decomposed into longitudinal and transverse components, denoted by $\mathbf{u}_l(t,\x)$ and $\mathbf{u}_t(t,\x)$, respectively. In addition, $c_{\lambda}$ and $D_{\lambda}$ represent the propagation speed and a parameter associated with the viscosity of the $\lambda$ branch, respectively. The subscript $\lambda$ labels both longitudinal and transverse displacement fields. From Eq.~\eqref{navier}, in the limit of extremely large viscosity, each Fourier component of $\mathbf{u}_{\lambda}(t,\x)$ satisfies $\Delta \mathbf{u}_{\lambda}(\,.\,,\x)=0$. For the transverse components of the sound wave in Eq.~\eqref{navier}, we can formulate a continuous field theory in $\mathbb{R}^d$.

Many authors have discussed a multivalley structure in the free energy of supercooled liquids and in the glassy state, connecting supercooling and vitrification with molecular-scale events \cite{goldstein, Francesco, Lubchenko, Langer}. In a supercooled liquid, the free energy landscape is a multidimensional potential-energy surface defined over the configuration space of particle coordinates.  To proceed, we extend this approach to coarse-grained fields to model the glass state of matter. Instead of using the space of all configurational degrees of freedom, which for $N$ particles is a $3N$-dimensional space, we define the free energy landscape in terms of local coarse-grained continuum fields. 
See, for example, Refs.~\cite{monasson,franz,mezarda,mezard,Rizzo}. 

{Our aim is to use a functional approach. Therefore, we resort to the Gibbs formulation of statistical mechanics, in which each state of the system is represented by a probability measure on phase space and observables are evaluated as expectation values.} Employing the functional-integral formulation of continuum field theory~\cite{livro5,justin1,roepstorff,justin}, we {propose a model where the emergent excitations of a structural glass raises from the many local minima of its free energy landscape}. In this framework, {we need first assume that} the relevant degrees of freedom are represented by a coarse-grained density field~$\Psi(\x)$. We decompose $\Psi(\x)$ as
\begin{equation}
\Psi(\x) = \Psi_{0} + \delta\Psi(\x), 
\end{equation}
where $\Psi_{0}$ denotes the average density of the glass. For notational convenience, we write the fluctuation field as $\delta\Psi(\x) = \varphi(\x)$.  To describe the metastable states, we employ a continuum field-theoretic model that extends beyond the Gaussian approximation and in which the static density-fluctuation field is coupled to a quenched disorder field. This disorder field accounts for environmental randomness; equivalently, the static density-fluctuation field is coupled to a random environment. 

It is important to emphasize that the metastable supercooled state is produced by a rapid quench of a thermodynamic control parameter, namely the temperature~$T = 1/\beta$. Since we do not {study the} dynamical model describing the evolution of the system from the high-temperature liquid phase down to the glass transition, the explicit temperature dependence of the effective field theory is inessential. Accordingly, we set~$\beta = 1$ without loss of generality. 

To establish our notation, we first consider a pure system, i.e., we neglect the effects of the disorder environment. In $\mathbb{R}^d$, we define random variables $\varphi(\x)$, distributed according to a probability measure $\mu(\varphi)$, where
\begin{equation}
    d \mu(\varphi) = Z^{-1} e^{-S(\varphi)} \prod_{\x} d\varphi(\x), 
\end{equation}
in which $Z$ is a normalization constant, i.e, the partition function and we define $ \prod_{\x} d\varphi(\x) \equiv [d\varphi]$, which is a formal measure over the space of all coarse-grained field configurations. {We define $S(\phi)$ as the action of the system.} The quantities associated with local observables are the $n$-point correlation functions, which we denote by $G(\x_{1},\dots,\x_{n})$. In Euclidean continuum field theory, we define the generating functional of these correlation functions, \(Z(j)\), as
\begin{equation}
Z(j)=\int [d\varphi]\exp \left(- S(\varphi) + \int d\x\, j(\x)\varphi(\x)\right),
\end{equation}
where $S(\varphi)$ is the action defining the model, and $j(x)$ is a fictitious external source. We write $Z(j)$ as a functional Taylor series,
\begin{align}
\frac{Z(j)}{Z(0)}=\sum_{n=0}^{\infty}\frac{1}{n!}
\int \prod_{i=1}^{n}d\x_{i}\, j(\x_{i}) \;
G(\x_{1}, \dots ,\x_{n}).
\end{align}
Taking functional derivatives with respect to the external source and then setting it to zero, one obtains the correlation functions $G(\x_{1}, \dots ,\x_{n})$ as  
\begin{align}
\label{eq:correlationfunction2}
G(\x_1,\cdots,\x_n) &= \frac{1}{Z(0)} \left.\frac{\delta^{k} Z(j)}{\delta j(\x_{1})...\delta j(\x_{n})}\right|_{j=0}  \nn \\[0.2true cm]
&= \frac{1}{Z(0)} \int [d\varphi]\, \varphi(\x_1)\cdots\varphi(\x_n) \, e^{-S(\varphi)} \nn \\[0.2true cm] 
&= \langle \varphi(\x_1)\cdots\varphi(\x_n)\rangle .
\end{align}
Notice that these $n$-point correlation functions are given by the sum of all diagrams with $n$ external legs, including disconnected diagrams. 

{Using the linked-cluster theorem, one can define the generating functional of connected correlation functions $G_c(x_1,\dots,x_n)$, or free energy\footnote{Our definition of the free energy differs by an overall minus sign from the conventional definition used in thermodynamics.}}:
\begin{align}
\hspace{-0.2cm}W(j) &= \ln Z(j) \nonumber \\[0.2true cm]
&=\sum_{m=0}^{\infty} \frac{1}{m!} 
\int \prod_{i=1}^{m} d\x_{i}\, j(\x_{i}) \; G^{(n)}_{c}(\x_{1},\dots,\x_{m}).
\end{align}
In our model, the action defining the pure system is given by:
\begin{equation} 
S(\varphi) = S_{0}(\varphi) + S_{I}(\varphi),
\label{eq:Apure}
\end{equation} 
with $S_{0}(\varphi)$ being the Gaussian contribution, given by
\begin{equation}
S_{0}(\varphi)= \frac{1}{2}\int d\x ´\;\varphi(\x) \left(-\triangle+m_{0}^{2} \right) \varphi(\x),
\label{eq:spe1}
\end{equation}
where $m_{0}^{2}$ is a spectral parameter, and $S_{I}(\varphi)$ is the self-interacting, non-Gaussian contribution:
\begin{equation}
S_{I}(\varphi) = \int d\x\,\left(\frac{\lambda_{0}}{4} \varphi^{4}(\x)+\frac{\rho_{0}}{6} \varphi^{6}(\x)\right).
\label{10}
\end{equation}

The Gaussian term \(S_{0}(\varphi)\) is required in a Euclidean formulation of field theory, as it regularizes the class of distributions that support the functional integration measure~\cite{Klauder}. {We also emphasize that, in principle, no ultraviolet cutoff parameter \(\Lambda\) is introduced to regularize potentially divergent expressions.} Instead, divergences will be treated by means of an analytic regularization scheme, both at the level of functional determinants and in the computation of \(n\)-point correlation functions.

The term $S_{I}(\varphi)$ must be a real polynomial bounded {from} below, i.e., it must be of even degree, with a positive coefficient for the highest-order term. As we will see shortly, the higher-order term $\rho_0 \varphi^{6}(\x)$ in $S_{I}(\varphi)$ is essential to ensure the existence of ground states for the coarse-grained field; for $\rho_{0}=0$, the disorder has a destructive effect if its intensity exceeds a certain value. We will show that our choice is the simplest one for the scenario of a random first-order phase transition. A scalar field-theory model with higher-order nonlinear terms was discussed in Ref.~\cite{adolfo}. We also note that, although it is possible~\cite{g1,g2} to couple an external source to the square of the field in the form $j(\x)\varphi^{2}(\x)$ and employ the composite-operator formalism of Ref.~\cite{CJT} to obtain nonperturbative results, here we use the traditional scenario with a local external source linearly coupled to the scalar field.

Next, we consider the coupling to a random environment. The generating functional for a given disorder realization $\eta(\x)$ is defined as
\begin{equation}
Z(\eta,j) = \int [d\varphi]\exp \left(- S_{II}(\varphi,\eta) + \int d\x\, j(\x)\varphi(\x)\right),
\label{Z-one}
\end{equation}
with
\begin{equation} 
S_{II}(\varphi,\eta) =  S(\varphi) - \frac{1}{2}
\int d\x \, \eta(\x) \, \varphi^{2}(\x),
\label{eq:Aprime} 
\end{equation}
where $S(\varphi)$ is the action of the pure system given in Eq.~\eqref{eq:Apure}. This is precisely the type of coupling that has been widely adopted in the literature to model a random environment. In most studies, the disorder is assumed to be Gaussian, with probability measure $[d\,\eta]P(\eta)$ and probability functional $P(\eta)$ given by
\begin{equation}
P(\eta) = p_{0}\exp\left(-\frac{1}{2\sigma^{2}}\int d\x\,\eta^{2}(\x)\right),
\end{equation}
where the parameter $\sigma$ characterizes the strength, or variance, of the disorder, and $p_{0}$ denotes a normalization constant. {Thus, the glassy phase in the class of materials considered here can be interpreted within a framework analogous to spin-glass physics.}

As mentioned above, $Z(\eta,j)$ in Eq.~\eqref{Z-one} is the generating functional of correlation functions for a single realization of the multiplicative disorder $\eta(\x)$, where $j(\x)$ is an arbitrary non-random external source introduced to generate correlation functions by functional differentiation. As in the pure system, we define a disorder functional $W(\eta, j)$ as follows:
\begin{align}
W(\eta,j) &= \ln Z(\eta,j).
\end{align}
Performing the disorder average, we get $\mathbb{E}\bigl[W(\eta,j)\bigr]$. Here, $\mathbb{E}\left[W(\eta,j)\right] =\mathbb{W}(j )$ means the disorder-average of the free energy for all multiplicative-disorder realizations, that is:
\begin{equation}
\mathbb{W}(j )= \int [d\,\eta]P(\eta)\,\ln Z(\eta,j),
\label{pro1}
\end{equation}
where $[d\,\eta]$ is a functional measure, i.e., a measure over all the disorder configurations, and $j(\x)$ is the external source \cite{brout}.
Using analytic techniques, one can show that the disorder-averaged free energy $\mathbb{W}(j)$ admits a series representation.

%
\section{The metastable phases in the glassy state of matter}\label{sec:ftg}
 
A significant technical challenge in disordered systems is the computation of the quenched average of the free energy. Several methods have been proposed in the literature, perhaps the most prominent being the replica trick~\cite{sg,livro1,livro3,nishimori}. In this approach, the average free energy in the presence of quenched disorder is obtained by considering the $k$-th power of the partition function, $Z^{\,k}=Z\times Z\times\cdots\times Z$. Here, $Z^k$ is initially interpreted as the partition function of a system composed of $k$ statistically independent copies of the original model. Using the identity $\ln Z = \lim_{k\to 0} (Z^k - 1)/k$, the quenched average is written as $\mathbb{E}[\ln Z] = \lim_{k\to 0} (Z_k - 1)/k$, where $Z_k = \mathbb{E}[Z^k]$ represents the partition function of the replicated system. In this limit, the disorder average couples the previously independent replicas, effectively leading to a zero-component field theory. Other powerful methods are also available for studying field theories in the presence of disorder, each offering distinct advantages in terms of rigor or physical intuition. Without attempting an exhaustive survey, we mention the supersymmetry method~\cite{Efetov1983} and the nonperturbative renormalization group~\cite{Dupuis:2020fhh}. In the present work, we use the distributional zeta-function method, which allows us to write the average free energy $\mathbb{W}(j)$ as
\begin{equation}\label{eq:completefreeenergy}
\mathbb{W}(j) = \sum_{k=1}^{\infty} \frac{(-1)^{k+1}}{k!\,k} \, \mathbb{E}\left[Z^{k}\right], 
\end{equation}
where the integer moment $\mathbb{E}\left[Z^{k}\right]$ is written as 
\begin{widetext}
\begin{equation}
\mathbb{E}\left[Z^{k}\right] = \int \prod_{i=1}^{k} [d\varphi_{i}^{(k)}] \, 
\exp \left(- S_{\rm eff}(\varphi^{(k)}_i) - \sum^k_{i=1} \int d\x \, j^{(k)}_i  \varphi^{(k)}_i  \right),
\end{equation}
where the action $S_{\rm eff}(\varphi^{(k)}_i)$ is given by  
\begin{align}
\label{eq:aa}
S_{\rm eff}(\varphi_{i}^{(k)}) &= \int d\x\,  \Biggl[
\sum_{i=1}^{k} \left[ 
\frac{1}{2}\varphi_{i}^{(k)}(\x)(-\Delta+m_{0}^{2})\varphi_{i}^{(k)}(\x) 
+\frac{\lambda_{0}}{4}\left(\varphi_{i}^{(k)}(\x)\right)^{4}
+\frac{\rho_{0}}{6}\left(\varphi_{i}^{(k)}(\x)\right)^{6}
\right] \nonumber \\
&-\frac{k\sigma^{2}}{4} 
\sum^k_{i,j=1} \left( \varphi_{i}^{(k)}(\x)\varphi_{j}^{(k)}(\x) \right)^{2}
\Biggr],
\end{align}
We now introduce a structure in function space for each term of the series, which leads to interesting consequences.
We stress that this choice reflects the idea that glass formation is associated with a fundamental change in the structure of the free energy landscape \cite{Ozawa}.
Our construction gives rise to random first-order transitions in the system.
We assume $j^{(k)}_{i}(\x)=j^{(k)}_{l}(\x)=j(\x)$ and make the following choice in the function space of the fields for each term of the series, i.e., for each integer moment $\mathbb{E}\left[Z^{k}\right]$ we have  $\varphi_{i}^{(k)}(\x) = \varphi_{j}^{(k)}(\x) \equiv \chi^{(k)}(\x)$. In the density functional theory of glass formation, replica off-diagonal density correlations are obtained by minimizing the free energy functional.
Our diagonal choice in function space disregards the off-diagonal correlations of density fluctuations in each moment of the partition function. Also, one can show that our choice minimizes the effective actions in each term of the series. As we will show, this approximation already yields two interesting results: {(i) with a particular choice in function space, we obtain random first-order transitions that model the glassy state of matter}, and (ii) we establish a connection between the boson peak and the presence of many metastable states and true ground states that characterize the average free energy of a each glassy state. Assuming the diagonal approximation { Eq. \eqref{eq:aa}}, one obtains:

\begin{equation}\label{eq:ezk11}
\mathbb{E}\left[Z^{k}\right] = {\cal N} \int [ d\chi^{(k)}] \, \exp \left( -S_{eff}(\chi^{(k)}) 
- k \int d\x\,  j(\x) \chi^{(k)}(\x)\right), 
\end{equation}
where
\begin{equation}\label{eq:effectivehamiltonian211}
S_{\rm eff}(\chi^{(k)}) = k \int d\x\, \left[
\frac{1}{2} \chi^{(k)}(\x) \left(-\Delta+m_{0}^{2}\right) \chi^{(k)}(\x) 
+\frac{1}{4}\left(\lambda_{0}-\frac{1}{2}k\sigma^{2}\right) \left(\chi^{(k)}(\x)\right)^{4}+\frac{\rho_{0}}{6}\left(\chi^{(k)}(\x)\right)^{6}
\right] .
\end{equation}
\end{widetext}

 In the series representation of the average free energy $\mathbb{W}(j)$, metastable states appear together with true ground states. {The series representation of the average free energy given by Eq.~(17), together with Eqs.~(20) and (21), models the many possible structurally disordered states that characterize the glassy phase}. For each term in the series representation of the average free energy, one can calculate the nucleation rate associated with the corresponding effective action; see, for example, Ref.~\cite{flores}. {Therefore, the dynamical slowdown is related to the transition rate out of the metastable states.}

The two-point correlation function can be obtained from the disorder-averaged generating functional of connected correlation functions, $\mathbb{W}(j)$. Information about the existence of distinct ground states is encoded in any static correlation function generated by $\mathbb{W}(j)$. We now connect this observation with thermodynamic amorphous order.
For $k$ larger than a critical value, new local ground states emerge.
For each integer moment of the partition function, the corresponding $k$-dependent ground states are located at
{\small \begin{equation}
\chi_{0}^{(k)} = \pm \left[\frac{1}{2\rho_{0}}\left(\frac{k\sigma^{2}}{2}-\lambda_{0}\right) 
+ \sqrt{\frac{1}{4\rho^2_{0}}\left(\frac{k\sigma^{2}}{2}-\lambda_{0}\right)^2 - \frac{m_0^2}{\rho_0}} \right]^{1/2}.
\end{equation}}
Therefore, to study fluctuations around the new ground states, we shift the fields as $\chi'^{(k)}(\x) = \chi^{(k)}(\x)-\chi_{0}^{(k)}$. The $k$-th free two-point correlation function $G_0^{(k)}(\x,\y)$ is then defined as
\begin{equation}
G_{0}^{(k)}(\x,\y)=\langle \chi'^{(k)}(\x)\chi'^{(k)}(\y)\rangle,
\end{equation}
satisfies
\begin{equation} 
\left(-\Delta+M_{0}^{2}\right)G_{0}^{(k)}(\x,\y)=\delta^{d}(\x-\y),
\end{equation} 
where $M_0^2$ is determined by the curvature of the shifted potential evaluated at the minimum.

To discuss the behavior of the two-point correlation function in the Gaussian approximation in a generic $d$-dimensional space, we consider the Laplace equation for $G_{0}^{(k)}(\x,\y)=\langle
\chi'^{(k)}(\x)\chi'^{(k)}(\y)\rangle$. In the high-dimensional limit, i.e., for $d\rightarrow\infty$, the Gaussian approximation becomes exact. We then have
\begin{equation}
\label{G0}
\left(-\Delta+M_{0}^{2}\right) G_{0}^{(k)}(\x,\y)=\delta^{d}(\x-\y), 
\end{equation}
where $M_{0}^{2}$ is an analytic function of the parameters of the model given by 
\begin{eqnarray}
(M_0(k))^{2}&=&-4m_0^2 + \frac{1}{\rho_0}\left(\frac{k\sigma^2}{2}-\lambda_0\right)^2 +\left(\frac{k\sigma^2}{2} - \lambda_0\right)\nonumber\\
&&\times \sqrt{\frac{1}{4\rho_0^2}\left(\frac{k\sigma^2}{2}-\lambda_0 \right)^2 - \frac{m_0^2}{\rho_0} }.
\end{eqnarray}

Using the Fourier representation of the two-point correlation function, one finds in $\mathbb{R}^d$ that
\begin{eqnarray}
G_{0}^{(k)}(\x-\y,M_{0})
&=&
\frac{1}{(2\pi)^{d/2}}
\left(\frac{M_{0}(k)}{|\x-\y|}\right)^{\frac{d-2}{2}} \nonumber\\
&&K_{\frac{d-2}{2}}\big(M_{0}(k)|\x-\y|\big),
\end{eqnarray}
where $K_{\nu}(z)$ is the modified Bessel function of the second kind. For $d=3$, we obtain
\begin{eqnarray}
G_{0}^{(k)}(\x-\y,M_{0})
=
\frac{1}{4\pi|\x-\y|}e^{-M_{0}(k)|\x-\y|}.
\end{eqnarray}
The above equation identifies a correlation length for each term in the
series. Thus, each moment of the partition function is associated with a
distinct correlation length. This provides a realization of thermodynamic
amorphous order that can propagate over intermediate length scales.

This is the endpoint of the static analysis. In the next section we use
this multivalley structure as the input for a coarse-grained elastic
description of the glassy phase. The key additional step will be to
interpret the distinct amorphous sectors as mesoscopic domains with
different effective elastic parameters, which naturally leads to a
random-coefficient wave equation for long-wavelength vibrations \cite{Ruocco,Arias}.

%
\section{Perturbation theory in glassy materials}\label{sec:ptdm}

To address the boson peak, we now study a nonstationary system evolving in time. We therefore consider elastic waves defined around the different ground states, namely the sound waves associated with each $k$-th moment in the series. These sound waves, denoted by $\mathbf{u}^{(k)}_\lambda(t,\x)$, where $\lambda = l,t$ labels the longitudinal and transverse polarizations, are described by the following wave equation:
\begin{equation}
	\left(\frac{1}{\bigl(c^{(k)}_\lambda\bigr)^2}\frac{\partial^2}{\partial t^2} - \Delta \right)\mathbf{u}^{(k)}_\lambda(t,\x)= 0,
    \label{wave-eq}
\end{equation}
where $c^{(k)}_l=u_{0}(k)$ and $c^{(k)}_t=u(k)$ are the longitudinal and transverse sound speeds, respectively. These velocities are effective long-wavelength quantities associated with the elastic response of a region of the glass whose static correlations are governed by the $k$-th distributional zeta-function sector. In this sense, the field theory determines the possible amorphous sectors and their static response, whereas the acoustic description provides the standard continuum limit for small-amplitude vibrations in an amorphous solid. In Eq.~\eqref{wave-eq}, we assume that the different polarization components are mutually decoupled. This assumption may not be strictly valid in a disordered medium, where distinct polarization components can couple to one another. Nevertheless, the prevailing view in the literature is that the boson peak originates predominantly from vibrational modes with transverse character. Consequently, as a first approximation, it is reasonable to restrict the analysis to transverse modes and neglect possible couplings to longitudinal modes. 

For notational simplicity, we denote $u^{(k)}_t(t,\bold{x})$ as $\phi^{(k)}(t,\x)$. The action corresponding to the wave equation for a transverse degree of freedom is given by 
\begin{equation}
S(\phi^{(k)})=
\frac{1}{2}\!\int\!dtd\x\, \biggl[\phi^{(k)}(t,\x)\biggl(\frac{1}{{u(k)}^2}\frac{\partial^{2}}{\partial t^{2}}
-\Delta\biggr)\phi^{(k)}(t,\x)\bigg].
\label{action34}
\end{equation}
where $u(k)=u(m_{0},\sigma^{2},k)$. From this action, one has that the (retarded) Green's function for the transverse waves in the $k$-th moment is given by  
\beq
G_k(\omega,\p)=\frac{u(k)^2}{\omega^2-u(k)^2\,p^2+i0^+}.
\eeq

To connect the multivalley structure with measurable spectral properties, we now introduce the
standard coarse-grained picture of a glass as a mosaic of mesoscopic domains.
We assume: (i) a harmonic regime described by linear elasticity; (ii) a separation of scales, with a typical
domain size $\ell$ such that all wavelengths of interest satisfy $\lambda\gg \ell$; (iii) a local
isotropic transverse response within each domain; and (iv) a stationary random-field description
of the spatial dependence of the elastic constants. A more detailed mathematical description of these assumptions is given in Appendix A.

Under these hypotheses, the glass is described by a transverse velocity field $u(\x)$ that is
approximately constant within each domain and takes values close to $u(k)$, depending on which
valley is locally realized. The physical statistics of the valleys are encoded in non-negative
weights $\pi_k$ (e.g., volume fractions) satisfying
\beq
\pi_k\ge 0,
\qquad
\sum_k \pi_k=1,
\label{eq:pik-def}
\eeq
which are independent modeling inputs. A minimal thermodynamic choice is to let $\pi_k$ be controlled by the valley free energy
density and by a degeneracy, or complexity, factor; however, the present derivation only requires
Eq.~(\ref{eq:pik-def}). {It is convenient to introduce the random field $\mu(\x)$ by defining}
\beq
\frac{1+\mu(\x)}{u_0^2}\equiv \frac{1}{u(\x)^2},
\qquad
\mu(\x)>-1,
\label{eq:mu-def-main}
\eeq
where $u_0$ is a reference effective velocity chosen such that $\lan\mu(\x)\ran=0$, i.e.
\beq
\frac{1}{u_0^2}=\left\langle \frac{1}{u(\x)^2}\right\rangle \simeq \sum_k \pi_k\,\frac{1}{u(k)^2}.
\label{eq:u0-harmonic-mean}
\eeq
The disorder strength is then measured by
\beq
\upsilon^2\equiv \lan \mu(\x)^2\ran \simeq \sum_k \pi_k\left(\frac{u_0^2}{u(k)^2}-1\right)^2.
\label{eq:upsilon-from-pi}
\eeq
We also assume short-range correlations with correlation length of order $\ell$,
\beq
\lan \mu(\x)\mu(\y)\ran = C(\x-\y),
\qquad
\widetilde C(\boldsymbol{q})=\int d\x\,e^{-i\boldsymbol{q}\cdot\x}C(\x).
\label{eq:Ccorr}
\eeq
In the lowest-frequency regime only the small-\(q\) behavior of \(\widetilde C(\boldsymbol{q})\) is relevant. In particular, for a smooth short-range covariance, \(\widetilde C(\boldsymbol{q})=\widetilde C(0)+O(q^2\ell^2)\). The commonly used white-noise model corresponds to the singular limit with the covariance proportional to the delta function; in practice the finite
$\ell$ provides the UV regularization.

With these definitions, the long-wavelength transverse excitations obey a random-coefficient
wave equation,
\beq
\biggl(\bigl(1+\mu(\x)\bigr)\frac{1}{u_{0}^{2}}\frac{\partial^{2}}{\partial t^{2}}-\Delta\biggr)
\psi(t,\x)=0,
\label{eq:rand-wave-main}
\eeq
which is the continuum implementation of spatially heterogeneous elasticity
\cite{sokolov,Gurarie1,Gurarie2,sheng}.

In Fourier space, the unperturbed Green's function is
\beq
G^{(0)}(\omega,\boldsymbol{k})=\frac{u_0^2}{\omega^{2}-u_0^2\, \boldsymbol{k}^2+i\delta},
\qquad \delta\to 0^+,
\label{eq:G0-main}
\eeq
and the disorder average can be performed perturbatively using a Dyson expansion.
Assuming an (approximately) Gaussian coarse-grained disorder with covariance (\ref{eq:Ccorr}),
the leading (Born) contribution to the averaged self-energy can be written schematically as
\beq
\Sigma_{\rm ave}(\omega)\;=\;\frac{\omega^4}{u_0^4}\int\!\frac{d^d q}{(2\pi)^d}\,
\widetilde{C}(\boldsymbol{q})\,G^{(0)}(\omega,\boldsymbol{q})\;+\;\cdots,
\label{eq:SigmaBorn-main}
\eeq
where $\widetilde{C}$ is the Fourier transform of $C$. In $d=3$, and for $\omega\ll \omega_D$ one
obtains the robust low-frequency scaling
\beq
\Re\,\Sigma_{\rm ave}(\omega)\sim \widetilde{C}(0)\,\omega^4,
\qquad
\Im\,\Sigma_{\rm ave}(\omega)\sim \widetilde{C}(0)\,\omega^5,
\label{eq:SigmaScaling-main}
\eeq
which corresponds to Rayleigh-type scattering of transverse modes.

The averaged Green's function can then be parametrized as
\beq
G^{(1)}(\omega,\boldsymbol{k})=\frac{1}{\omega^2-u_0^2 k^2+\Sigma_{\rm ave}(\omega)/u_0^2},
\label{eq:G1-main}
\eeq
and the spectral density (normalized as in Appendix~\ref{sec:fel}) is
\beq
g(\omega)=-\frac{\omega}{\pi^3 k_D^3}\int_0^{k_D} dk\,k^2\,\Im\bigl[G^{(1)}(\omega,k)\bigr].
\label{eq:dos-main}
\eeq
Using (\ref{eq:SigmaScaling-main}) one obtains, in $d=3$, the low-frequency expansion
\beq
g(\omega)=\frac{\omega^2}{2\pi^2 u_0^3 k_D^3}\left[1+\mathcal{A}\,\omega^2+\cdots\right],
\qquad
\mathcal{A}\propto \widetilde{C}(0),
\label{eq:dos-omega4-main}
\eeq
i.e. an excess contribution $\delta g(\omega)\propto \omega^4$ beyond the Debye law.
Therefore, $g(\omega)/\omega^2$ increases with $\omega^2$ at low frequency, which is the
characteristic onset of the boson-peak anomaly.

For an explicit evaluation of the prefactors (including the Debye cutoff and the
white-noise approximation), one may use the calculation in Appendix~\ref{sec:fel}, which leads to
Eq. (B21) for $\Sigma_{\rm ave}(\omega)$ and reproduces the $\omega^4$ correction to the
spectral density. The emergence of an actual maximum (a true peak) at intermediate
frequencies requires going beyond the strict low-$\omega$ Born regime, e.g. incorporating the
finite correlation length $\ell$ through $\widetilde{C}(\boldsymbol{q})$ and/or using a
self-consistent (effective-medium) resummation of the Dyson series.

\section{Conclusions}\label{sec:conclusions}

{The transition of a supercooled liquid to a glassy state is a complex phenomenon characterized by several physical features. Here, we highlight four of them: (i) an enormous increase in viscosity, (ii) a dramatic dynamical slowdown without significant changes in the microscopic structure of the system, (iii) fragility, and (iv) dynamical heterogeneity, namely the coexistence of liquid-like and solid-like behavior during the transition from the supercooled liquid to the glassy state.}

{A theoretical description of items (i), (ii), and (iii) requires addressing the complex problem of modeling the microscopic degrees of freedom involved in the glass transition and their interactions. There are two main approaches to modeling vitrification without crystallization in a supercooled liquid: one invokes thermodynamics, whereas the other regards the transition as a dynamical phenomenon. We show that feature (iv) allows us to implement an effective field theory, based on coarse-grained fields with randomness, and thereby describe the glassy state of matter as a thermodynamic phenomenon.}.

Using the formalism of a continuous field theory in Euclidean space, we define static density-fluctuation fields to model the glassy state of matter. The effective description is based on a Landau-Ginzburg model with quenched randomness, in which the coarse-grained fields are coupled to multiplicative quenched disorder. By performing disorder averages, we show that the average free energy of the system admits a functional series representation. This representation describes the multivalley structure of the average free energy of the system and characterizes the glassy state. {This construction realizes what is known in the literature as random first-order transition.} 

As we discussed, the lack of {crystalline} periodicity raises fundamental questions concerning the spectral properties of the glassy state. In particular, glasses exhibit an excess vibrational density of states over the Debye prediction. To study emergent excitations in glasses, we use the fact that the average free energy of a single glassy state contains many local minima. A key consequence is the emergence of elastic behavior in the glassy state, which makes it possible to formulate hyperbolic differential equations with random coefficients. In Refs.~\cite{boson,rizzo1}, the boson peak was related to saddle points of the free energy landscape of the glass. Here, we instead explore the connection between the many local minima that characterize the average free energy of a single glassy state and the use of hyperbolic differential equations with random coefficients to describe emergent excitations in amorphous materials. {It should be emphasized that the spectral properties of a low-temperature liquid above the crystallization temperature require a different formalism, because metastable states are absent, see, for example, Ref.~\cite{liquids}}.

We now call attention to the interpretation of the series obtained for the average free energy using the distributional zeta-function method. Within the diagonal approximation in function space, the series becomes a superposition of different moments of the partition function, each exhibiting distinct ground states. Regarding the question of whether vitrification is essentially thermodynamic, our results show that it can be described as a nonconventional thermodynamic transition involving a plethora of metastable and ground states. Furthermore, our results show that random differential equations can be used to describe the propagation of excitations in structural glasses, as emphasized in the literature. In particular, we find a contribution to the spectral density $D(\omega)$ proportional to $\omega^{4}$.  

The natural continuations of this work is to study the thermodynamic properties of the low-temperature liquids and glassy state of matter. Using spectral methods, one can discuss the entropy of those materials described by the random-field model, i.e., the low-temperature liquid, as well the glassy state of matter, model by a field theory of density fluctuations with multiplicative disorder~\cite{prep}. Our aim is to discuss the Kauzmann paradox \cite{kauzmann1, stillinger1, kivelson, stillinger2, hajimet, speedy, rebecca} using functional methods and disorder fields.

\begin{acknowledgments} 
We would like to thank G. Heymans, C. Farina and B. F. Svaiter for useful discussions. {This work was partially supported by Conselho Nacional de Desenvolvimento Cient\'{\i}fico e Tecnol\'{o}gico (CNPq) grants nos. 141376/2023-6 (IPF), 303436/2015-8 (NFS), 313254/2025-7 (GK), 307626/2022-9 (AMSM), Funda\c{c}\~{a}o de Amparo \`{a} Pesquisa do Estado de S\~{a}o Paulo (FAPESP), grant nos. 2018/25225-9 (GK), 2021/14335-0 (GK)}, and Coordena\c{c}\~{a}o de Aperfei\c{c}oamento de Pessoal de N\'{\i}vel Superior (CAPES), grant no. 88887.820425/2023-00 (MMB). 

\end{acknowledgments}

\appendix

\section{From the multivalley landscape to the random wave equation}

To pass from the static free energy landscape to a dynamical description of elastic excitations, we introduce three hypotheses. These hypotheses should be understood as physical coarse-graining assumptions, not as exact consequences of the distributional zeta-function series. At mesoscopic scales, the physical glass in $d=3$ spatial dimensions can be described as a stationary mosaic of amorphous regions $\{\Omega_\alpha\}$ satisfying the following conditions:
\begin{enumerate}
\item[(i)] each region $\Omega_\alpha$ is locally described by one of the amorphous sectors labeled by $k_\alpha\in\{1,2,\ldots,N\}$;
\item[(ii)] the typical linear size $\ell$ of such regions satisfies $a\ll \ell\ll L$, where $a$ is the microscopic scale and $L$ is the macroscopic sample size;
\item[(iii)] the field $K(\x)=k_\alpha$ for $\x\in\Omega_\alpha$ is statistically homogeneous, with volume fractions
\begin{equation}\label{eq:pik-def1}
\pi_k\ge 0,
\qquad
\sum_{k=1}^{N}\pi_k=1.
\end{equation}
\end{enumerate}
These hypotheses provide a spatial realization of the standard glassy picture, in which a frozen amorphous solid samples different metastable configurations across mesoscopic regions. The distributional zeta-function construction identifies a family of amorphous sectors together with their static response functions. The non-negative weights $\pi_k$ represent the physical volume fractions of these sectors for a given glass preparation. These fractions may depend on the quench protocol, the sector free energies, and configurational degeneracies, but a detailed model for them is not required for the low-frequency calculation below.

In each mesoscopic region $\Omega_\alpha$ described by a sector $k_\alpha$, the small-amplitude transverse excitations are governed, at wavelengths large compared with the microscopic scale, by an effective elastic wave equation
\begin{equation}\label{eq:wavek}
\left(
\frac{1}{\bigl(u(k_\alpha)\bigr)^2}
\frac{\partial^2}{\partial t^2}
-
\Delta
\right)
\phi^{(k_\alpha)}(t,\x)=0,
\qquad
\x\in\Omega_\alpha.
\end{equation}

The transverse sound speed $u(k)$ is a sector-dependent effective parameter determined by the local amorphous structure and constrained by the corresponding static response scale $\xi_k$. This is the standard continuum-elasticity step: each frozen amorphous region is characterized by local elastic constants and, consequently, by local sound speeds. The distributional zeta-function formalism provides a field-theoretic basis for the existence of multiple amorphous sectors with distinct static correlation lengths and susceptibilities. Since the boson peak is commonly associated with transverse-dominated low-frequency modes, we retain only the transverse scalar reduction. Possible longitudinal-transverse mixing constitutes a higher-order refinement of the effective theory.

The above discussion implies that the transverse field over the whole sample obeys a random-coefficient wave equation. Defining the local inverse squared speed
\begin{equation}\label{eq:mx}
m(\x)\equiv \frac{1}{u^2(\x)}
=\sum_\alpha \frac{1}{u^2(k_\alpha)}\,\mathbf{1}_{\Omega_\alpha}(\x),
\end{equation}
where $\mathbf{1}_{\Omega_\alpha}$ is the indicator function of the region $\Omega_\alpha$, the effective equation reads
\begin{equation}\label{eq:fullwave}
\left(m(\x)\frac{\partial^2}{\partial t^2}-\Delta\right)\psi(t,\x)=0.
\end{equation}
This is the same class of random wave equations used in phenomenological approaches to the boson-peak, but here the origin of the random coefficients is tied to the multivalley structure of the distributional zeta-function average free energy.

To proceed analytically with Eq.~\eqref{eq:fullwave}, we decompose the local inverse squared speed into its mean and fluctuation:
\begin{equation}
 m(\x)=\bar m+\delta m(\x),
\qquad
\bar m=\lan m(\x)\ran.
\end{equation}
The brackets denote an average over the stationary mosaic ensemble. We define the reference velocity $\bar u$ and the dimensionless fluctuation $\mu(\x)$ by
\begin{equation}\label{eq:mudef}
\bar u=\bar m^{-1/2},
\qquad
\mu(\x)=\frac{\delta m(\x)}{\bar m},
\qquad
\lan\mu(\x)\ran=0.
\end{equation}
Equivalently,
\begin{equation}\label{eq:u0-harmonic}
\frac1{\bar u^2}
=\left\langle\frac1{u^2(\x)}\right\rangle
\simeq
\sum_{k=1}^{N}\pi_k\frac1{u^2(k)}.
\end{equation}
The wave equation becomes
\begin{equation}\label{eq:randwave}
\left(\bigl(1+\mu(\x)\bigr)
\frac{1}{\bar u^2}
\frac{\partial^2}{\partial t^2}
-
\Delta
\right)
\psi(t,\x)=0.
\end{equation}
This is the minimal random-wave model for transverse excitations in a heterogeneous amorphous solid. At wavelengths large compared with the mosaic scale, the fluctuation field $\mu(\x)$ may be approximated by a stationary, short-range correlated random field with
\begin{equation}\label{eq:mucorr}
\lan\mu(\x)\ran=0,
\qquad
\lan\mu(\x)\mu(\y)\ran=C(\x-\y),
\end{equation}
where $C(\r)$ decays on a length scale of order $\ell$. Its Fourier transform satisfies
\begin{equation}\label{eq:Ctilde}
\widetilde C(\mathbf{q})\ge 0,
\qquad
\widetilde C(\mathbf{q})=\widetilde C(0)+\mathcal{O}(q^2\ell^2)
\quad (q\ell\ll 1).
\end{equation}

For weak effective disorder, the leading correction to the spectral density is controlled by the covariance $C$, independently of the detailed non-Gaussian structure of the mosaic. The Gaussian approximation should therefore be regarded as a convenient weak-disorder closure, rather than as an essential microscopic assumption. At long wavelengths, the wave samples many mesoscopic regions, so the leading Born correction depends only on the covariance of the elastic fluctuations, while higher cumulants contribute only at higher orders. A useful measure of the local variance is
\begin{equation}\label{eq:localvar}
\upsilon_0^2
\equiv
\lan\mu(\x)^2\ran
\simeq
\sum_{k=1}^{N}\pi_k
\left(
\frac{\bar u^2}{u^2(k)}-1
\right)^2,
\end{equation}
while the quantity entering the low-frequency scattering amplitude is the integrated covariance
\begin{equation}\label{eq:C0integrated}
\widetilde C(0)=\int d\x\, C(\x)
\sim
\upsilon_0^2\ell^3,
\end{equation}
up to a shape factor depending on the detailed correlation function. The white-noise model used in many calculations corresponds to the formal limit
\begin{equation}\label{eq:white-noise-limit}
C(\x)\to \upsilon^2\delta^{(3)}(\x),
\qquad
\widetilde C(\mathbf{q})=\upsilon^2,
\end{equation}
with an implicit ultraviolet cutoff. In an amorphous solid the finite length $\ell$ provides the physical cutoff.

Thus, the effective disorder strength is not arbitrary in the sense of a purely phenomenological model: it is constrained by the sector-dependent responses generated by the distributional zeta-function construction and by the preparation-dependent fractions $\pi_k$. At the same time, it is not determined solely by the local potential parameters; it also depends on how the glass samples the amorphous sectors in space.

\section{Spectral density in glassy materials}\label{sec:fel}

Let us discuss a scalar field $\psi(t,\x)$ defined in a $d+1$ dimensional space-time that satisfies the random wave  equation
\begin{equation}
\biggl(\bigl(1+\mu(\x)\bigr)\frac{1}{u_{0}^{2}}\frac{\partial^{2}}{\partial t^{2}}-\Delta\biggr)\psi(t,\x)=0.
\end{equation} 

We have shown that the low-frequency excitations of the amorphous material are described by equations of motion involving macroscopic properties of the system, such as elastic constants. Disorder induces spatial fluctuations in these quantities.
The field $\mu(\x)$ encodes the resulting randomness in the amorphous solid and is treated here as annealed disorder. Let us define the Fourier transforms of $\psi(t,\x)$ and also $\mu(\x)$. We have 
\begin{equation}
\psi(t,\x)=\frac{1}{2\pi}\int\,d\omega\,e^{-i\omega t}\psi(\omega,\x),
\end{equation}
and also
\begin{equation}
\psi(t,\x)=\frac{1}{(2\pi)^{d}}\int\,d\bold{k}\,e^{-i\bold{k}\cdot\x}\psi(t,\bold{k}).
\end{equation}
In addition, we also define the Fourier transform of the stationary random function. We have
\begin{equation}
\mu(\x)=\frac{1}{(2\pi)^{d}}\int\,d\bold{k}\,e^{-i\bold{k}\cdot\x}\mu(\bold{k}).
\end{equation}
By substituting the Fourier representations of $\psi(t,\x)$ and $\mu(\x)$ into the random wave equation, one finds that the Fourier components $\psi(\omega,\bold{k})$ of the field satisfy 
\begin{equation}
\int d\bold{k}'\bigl(L_{0}(\bold{k},\bold{k}')+L_{1}(\bold{k},\bold{k}')\bigr)\psi(\omega,\bold{k}')=0,
\end{equation}
where $L_{0}$ and  $L_{1}$ are non-random and random matrices respectively with elements
\begin{equation}
L_{0}(\bold{k},\bold{k}')=\Biggl(\frac{\omega^{2}}{u_{0}^{2}}-\bold{k'}^{2}\Biggr)\delta(\bold{k}-\bold{k}')
\end{equation}
and 
\begin{equation}
L_{1}(\bold{k},\bold{k}')=\frac{1}{(2\pi)^{d}}\biggl(\frac{\omega^{2}}{u_{0}^{2}}\mu(\bold{k}-\bold{k}')\biggr).
\end{equation}

The $L_0$ and $L_1$ are integral convolution operators. The random wave equation can be written in the form
\begin{equation}
\bigl(L_{0}+L_{1}\bigr)\psi(\omega,\cdot)=0.
\end{equation}
Next, we define the operator-valued Green's function $G$:
\begin{equation}
G=\bigl(L_{0}+L_{1}\bigr)^{-1}.
\end{equation}
Using the Dyson perturbative expansion, where $G^{(0)}=L^{-1}_{0}$ is the unperturbed operator valued Green's function, we can write
\begin{equation}
G=G^{(0)}-G^{(0)}L_{1}G^{(0)}+G^{(0)}L_{1}G^{(0)}L_{1}G^{(0)}+...
\end{equation}
Let us define the self-energy $\Sigma$ given by
\begin{equation}
\Sigma=L_{1}-L_{1}G^{(0)}L_{1}+...
\end{equation}
The Fourier components of the unperturbed operator value Green's function $G^{(0)}$ are given by
\begin{align}
G^{(0)}(\omega,\bold{k})=\frac{u^2_0}{\omega^{2}-u_0^2\, \bold{k}^2+i\delta}.
\end{align}
To proceed, let us write the perturbative series in $\x$ space, up to the second order of randomness:
\begin{widetext} 
\begin{align}
\label{randg}
G(\omega,\x,\x')&\approx G^{(0)}(\omega,\x,\x') 
-  \omega^2\int d\mathbf{x}_1\, G^{(0)}(\omega,\x,\x_1)\, \mu(\x_1) \,G^{(0)}(\omega,\x_1,\x') 
\nonumber \\
&+  \omega^4\int d\x_1 \int d\x_2 \, G^{(0)}(\omega,\x,\x_2) \, \mu(\x_2) \, G^{(0)}(\omega,\x_2,\x_1)\,\mu(\x_1)\,G^{(0)}(\omega,\x_1,\x').
\end{align}
This expression shows that spatial correlations are modified by the random field. Averaging $G(\omega, \x, \x')$ over the realizations of $\mu(\x)$ yields the effective contribution of disorder to the self-energy. We define the disorder-averaged Green function $G^{(1)}(\omega, \x, \x')$ as
\begin{equation}
    G^{(1)}(\omega, \x, \x') = \int [d\mu] \, P(\mu) \, G(\omega, \x, \x'),
\end{equation}

\noindent where $P(\mu)$ is the density probability functional of the disorder field $\mu(\boldsymbol{x})$. Considering a Gaussian probability functional , 
\begin{equation}
    P(\mu) = P_0\exp{\left(-\frac{1}{2\upsilon^2} \int d\x \, \mu^2(\x)\right)},
\end{equation}
and since $G^{(0)}(\omega, \bold{x}, \bold{x}')$ does not depend on the disorder field $\mu(\bold{x})$, its average over the random process is simply $G^{(0)}(\omega, \bold{x}, \bold{x}')$ itself and, therefore, the average $G^{(1)}(\omega, \mathbf{x}, \mathbf{x}')$ is written as
\begin{equation}
G^{(1)}(\omega, \x, \x') = G^{(0)}(\omega, \x, \x')
- G^{(0)}(\omega, \bold{x, 0})\left(\int d\mu \, P(\mu) \Sigma(\omega, \bold{x}, \bold{x}', \mu) \right)G^{(0)}(\omega, \bold{x}',\bold{0}),
\end{equation}
\end{widetext}
where $\Sigma_{ave}(\omega, \x, \x')$ is the average of the self-energy: 
\begin{equation} 
\Sigma_{ave}(\omega, \x, \x') = \int d\mu \, P(\mu) \Sigma(\omega, \bold{x}, \bold{x}', \mu)
\end{equation} 
The procedure to calculate $\Sigma_{eff}(\omega, \bold{p})$ is very similar to the one presented in Ref. \cite{Arias}. The resulting value for $\Sigma_{eff}(\omega)$ is:
\begin{eqnarray}
    \Sigma_{ave}(\omega) &=& -\frac{\upsilon^2 \omega^4}{(2\pi)^{d-1}k_D^3} \int d \bold{k} \, G^{(0)}(\omega, \bold{k}) \nonumber\\
    &=& \frac{u_0^2 \, \upsilon^2 \, \omega^4}{(2\pi)^{d-1}k_D^3} \int d\bold{k}\, \frac{1}{u_0^2\, \bold{k^2} - \omega^2}. 
\end{eqnarray}

The above integration is solved for $d=3$. Thus,
\begin{equation}
    \Sigma_{ave}(\omega) = \frac{u^2_0\upsilon^2 \omega^4}{(2\pi)^3 k_D^3}\int_0^{k_D} dk \frac{k^2}{u^2_0 \,k^2-\omega^2 + i\delta}, \quad \delta \rightarrow 0
\end{equation}

\noindent where $k=|\bold{k}|$ and
 $k_D$ is the Debye momentum. The integration can be carried out, leading to the effective induced coupling:
\begin{equation}
     \Sigma_{ave}(\omega) = \frac{\upsilon^2 \omega^4 }{2\pi^2 u_0 k_D^3} \omega_D - \frac{\upsilon^2 \omega^5}{4\pi u_0 k_D^3} \ln{\left|\frac{1 + \frac{\omega_D}{\omega}}{1 - \frac{\omega_D}{\omega}}\right|} +  i\frac{\upsilon^2\omega^5}{4\pi u_0 k_D^3}.
\end{equation}

Taking into account only the low-frequencies regime ($\omega \ll\omega_D$), the average of the self-energy is given by:
\begin{equation}
    \Sigma_{ave}(\omega) \approx \frac{\upsilon^2 \omega^4}{2\pi^2 u_0 k_D^3 } \omega_D + i\frac{\upsilon^2\omega^5}{4\pi u_0 k_D^3}.
\end{equation}\label{self2}

The Green's function $G^{(1)}(\omega, \mathbf{k})$ is related to the average of the self energy in the following way, see reference \cite{Arias}:
\begin{equation}
    G^{(1)}(\omega, \mathbf{k}) = \frac{1}{\omega^2 - u_0^2k^2 + \frac{\Sigma_{ave}(\omega)}{u_0^2}}.
\end{equation}\label{g1}

From $G^{(1)}(\omega, \mathbf{k})$, we calculate the spectral density of the glass, which is given by:
\begin{equation}
    g(\omega) = -\frac{\omega}{\pi^3 k_D^3}\int_0^{k_D} dk \, k^2\, \mathrm{Im}[ G^{(1)}(\omega, k)]. 
\end{equation}

The density of states of the amorphous solid in the low-frequency regime is given by:
\begin{equation}
    g(\omega) \approx \frac{\omega^2}{2\pi^2 u_0^3 k_D^3}\left(1 + \frac{\upsilon^2 \omega_D}{4\pi^2 u_0 k_D^3 u_0^2} \omega^2 \right).
\end{equation}
\vspace{0.1cm}

This is the expected low-frequency dependence of the spectral density of an amorphous solid and is associated with the boson-peak phenomenon. The prefactor of the $\omega^4$ contribution has been discussed extensively in the literature. The disorder strength encodes both material-dependent properties and preparation-dependent features of the glassy state. Thus, for a given glass former, it can be used to model the dependence of the glassy state on the parent temperature.

\end{document}